\documentclass[twocolumn,showpacs,preprintnumbers,amsmath,amssymb]{revtex4}
\usepackage{graphicx}% Include figure files
\usepackage{dcolumn}% Align table columns on decimal point
\usepackage{bm}% bold math
\begin{document}

\title{Comment on \lq\lq Maximal planar networks with large clustering coefficient and power-law degree distribution \rq\rq}

\author{Zhi-Xi Wu \footnote {Electronic address: wupiao2004@yahoo.com.cn}}
\author{Xin-Jian Xu}
\author{Ying-Hai Wang \footnote {Electronic address: yhwang@lzu.edu.cn}}
\affiliation{Institute of Theoretical Physics, Lanzhou University, Lanzhou
Gansu 730000, China}

%\date\today
\begin{abstract}
This Comment corrects the error which appeared in the calculation of the degree
distribution of random apollonian networks. As a result, the expression of
$P(k)$, which gives the probability that a randomly selected node has exactly
$k$ edges, has the form $P(k)\propto 1/[k(k+1)(k+2)]$.
\end{abstract}
\pacs{89.75.Hc} \maketitle

In a recent article \cite{Zhou}, Zhou \emph{et.al.} discussed a class of
networks, called random apollonian networks (RAN), which display scale-free
degree distribution, very large clustering coefficient and very small average
path length. The simultaneous small-world and scale-free properties can mimic
well many real-life network systems.

However the analysis errs in calculating the degree distribution function. The
rate equation they used to derive the expression of  $P(k)$ has the form:
\begin{equation}
n(N+1,k+1)=n(N,k)\frac{k}{N_{\triangle}}+n(N,k+1)(1-\frac{k+1}{N_{\triangle}}),\label{eq1}
\end{equation}
where $n(N,k)$ denotes the number of nodes with degree $k$ when $N$ nodes are
present in the network, and $N_{\triangle}$, the total number of triangles.
This rate equation describes the time evolution of the number of nodes with
degree $k+1$. The first term accounts for the process in which a site with $k$
links is connected to the new site, leading to a gain in the number of sites
with $k+1$ links. This happens with probability $\frac{k}{N_{\triangle}}$. The
second term on the right-hand side of Eq. (\ref{eq1}) accounts for a
corresponding role (loss). Using the asymptotic limit $n(N,k)=NP(k)$ and
$P(k+1)-P(k)=dP/dk$, they obtained the form of the degree distribution
function: $P(k)\propto k^{-\gamma}$ with $\gamma=(3N+5)/N$. However, as we will
state below, that $\frac{k}{N_{\triangle}}$ does not give the right probability
of nodes with degree $k$ gaining a new link. Thus their result for $P(k)$ is
incorrect.

The growth of a RAN is performed by randomly selecting a triangle in the
network, and then a new node connects to the three vertices of the triangle
(see Ref. \cite{Zhou} for details). This means that the correlation among the
nodes is very strong, i.e., after a new link added to a node, the remaining two
links are constrained to link to it's two neighbors. Thus, on the one hand
$\frac{k}{N_{\triangle}}$ would include the probability of the evolution of
other nodes whose degree is not equal to $k$; on the other hand, the evolution
of the neighbors (except for those neighbors with degree $k$) of a node (this
node has degree $k$) would effect it's evolution in the reverse, which could
lead to the information missing for the evolution of $n(N,k)$. The same
analysis is also suitable for the second term of the  Eq. (\ref{eq1}). In the
following, we give out the right form of the expression of $P(k)$.

\begin{figure}
\includegraphics*[width=8.4cm]{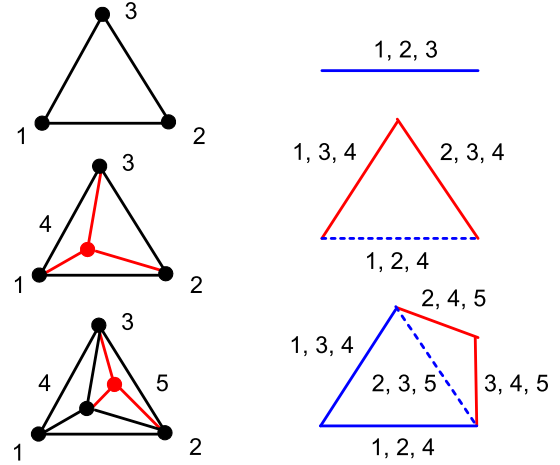}
\caption{Schematic illustration of the growth of RAN (left column) and AN
(right column) for $t=0, 1$ and $2$ (from top to bottom). Each triangle in the
RAN is represented by an edge in the AN. That a node added to the RAN
corresponds to an edge is replaced by a triangle in the AN. Left: the red node
and lines denote the newly added node and links. Right: the dash blue line
depict the random selected edge (to update) and the red lines denote the two
newly added edges after each generation. Note that the information included in
the edges are also renewed at the same time (the coordinates of nodes in the
RAN denoted by the edges in the AN).} \label {fig1}
\end{figure}

\begin{figure}
\includegraphics*[width=8.4cm]{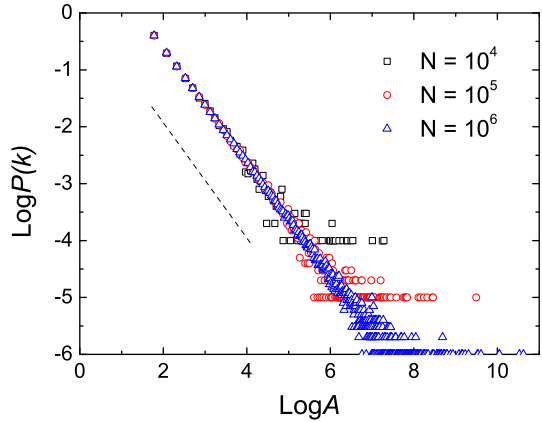}
\caption{The distribution of $P(k)$ as a function of A, where A=$k(k+1)(k+2)$,
for different network sizes. The dash line is a guide to the eyes with slope
-1.0.} \label {fig2}
\end{figure}

Instead of $n(N,k)$, let us investigate the time evolution of $p(k,t_i,t)$, the
probability that at time $t$ a node $i$ introduced at time $t_i$ has a degree
$k$. This \lq\lq node-based-method \rq\rq can overcome the strong node-node
correlation arising in the \lq\lq group-based-method \rq\rq. Indeed, the growth
of the RAN can be understood by the evolution of an adjacent network (for
convenience, we call it AN): each triangle in the RAN is represented by an
edge; whenever two triangles share a edge, the corresponding two edges are
connected in the AN; That a new node is added to the RAN (leading to two
triangles increasing ) corresponds to a randomly selected edge is replaced by a
new triangle  in the AN (this leads to two edges increasing) (see figure 1).
Note that each edge in the AN represents three nodes in the original RAN. When
a new node enters the system, the degree of node $i$ increases by $1$ with a
probability dependent on the emergence times of this node on all the edges of
the AN, which is equal to the number of triangles containing the $i$th node in
the RAN. Let $N_{\triangle}^i$ denote this number. For large time limit,
$N_{\triangle}^i$ is equal to the degree of the $i$th node:
$N_{\triangle}^i=k_{i}$.

Thus, for $i$th node, the probability of gaining a new link (or the emergence
times increasing by 1) is $k_i/2t$, where $2t$ denotes the total edges in the
AN at time $t$ (note that each updating step would give rise to two extra edges
increasing). Consequently the master equation governing $p(k,t_i,t)$ for the
RAN has the form \cite{Albert, Dorogovtsev},
\begin{equation}
p(k,t_i,t+1)=\frac{k-1}{2t}p(k-1,t_i,t)+(1-\frac{k}{2t})p(k,t_i,t)\label{eq2}
\end{equation}
with the initial condition $p(k, t_i=0,1, t=1 )=\delta_{k,1}$ and the boundary
one $p(k,t,t)=\delta_{k,1}$. The degree distribution can be obtained as
\cite{Albert}:
\begin{equation}
P(k)=\lim_{t \rightarrow \infty}(\sum_{t_i}p(k,t_i,t))/t.
\end{equation}
Using Eq. (\ref{eq2}) and the expression $p(k)-p(k-1)=dp/dk$, one can get that
$P(k)$ is the solution of the recursive equation: $P(k)=P(k-1)(k-1)/(k+2)$ for
$k\geq m+1$, and $P(m)=2/(m+2)$ for $k=m$, where $m$ is the degree of a node at
the time it enters the sytem (in the present case $m=3$). Solving for $P(k)$
gives that
\begin{equation}
P(k)=\frac{2m(m+1)}{k(k+1)(k+2)} \label{eq3}.
\end{equation}

In Fig. \ref{fig2}, the simulation results of $P(k)$ for different network
sizes $N=10^4, 10^5$ and $10^6$ are plotted as a function of $k(k+1)(k+2)$, the
best fit line gives out that the decay exponent is $-1.002\pm0.003$, which is
in well agreement with the analytic result Eq. (\ref {eq3}).


\begin{references}
\bibitem{Zhou}
T. Zhou, G. Yan and B.H. Wang, Phys. Rev. E \textbf{71}, 046141 (2005).

\bibitem{Albert}
R. Albert and A.-L. Barab\'asi, Rev. Mod. Phys. \textbf{74}, 47 (2002).

\bibitem{Dorogovtsev}
S.N. Dorogovtsev and J.F.F. Mendes, Adv. Phys. \textbf{51}, 1079 (2002).
\end{references}
\end{document}